\documentclass[aps,prl,reprint,superscriptaddress,amsmath,amssymb,floatfix]{revtex4-2}
\usepackage[utf8]{inputenc}
\usepackage{CJKutf8}
\usepackage{graphicx}
\usepackage{bm}
\usepackage{dsfont}
\usepackage{hyperref}
\hypersetup{colorlinks=true,linkcolor=blue,citecolor=blue,urlcolor=blue}
\begin{document}
\title{Universal Two-Excitation Scattering in Two-Dimensional Subwavelength Atomic Arrays}
\author{Yidan Wang (\begin{CJK*}{UTF8}{gbsn}王艺丹\end{CJK*})}
\affiliation{Department of Physics, Harvard University, Cambridge, Massachusetts 02138, USA}
\author{Oriol Rubies-Bigorda}
\affiliation{Physics Department, Massachusetts Institute of Technology, Cambridge, Massachusetts 02139, USA}
\affiliation{Department of Physics, Harvard University, Cambridge, Massachusetts 02138, USA}
\author{Valentin Walther}
\affiliation{Department of Physics and Astronomy, Purdue University, West Lafayette, Indiana 47907, USA}
\affiliation{Department of Chemistry, Purdue University, West Lafayette, Indiana 47907, USA}
\author{Susanne F. Yelin}
\affiliation{Department of Physics, Harvard University, Cambridge, Massachusetts 02138, USA}

\begin{abstract}
Subwavelength atomic arrays are a leading platform for engineering strong light-matter interactions, presenting exciting opportunities for quantum science. However, a full understanding of their multi-excitation dynamics remains a significant challenge. In this Letter, we uncover a remarkable universal phenomenon that emerges in these arrays. Using scattering theory to analyze two-excitation interactions, we reveal a profound simplification near critical points of the collective atomic excitation band structure, determined solely from single-excitation properties. At these critical points, which emerge as band edges and saddle points (Van Hove singularities) in the two-excitation phase space, scattering becomes universal and the full two-excitation scattering matrix decomposes into a block-diagonal form. Remarkably, all scattering processes involving the photon field are completely suppressed, resulting in the perfect isolation of a unitary, nonlinear interaction channel between collective dark spin waves. Our findings provide exact analytical insights into few-body nonlinearities and establish a universal framework for analyzing complex scattering phenomena in ordered atomic systems.
\end{abstract}
\maketitle

The pursuit of universal principles that distill simple, predictable behaviors from complex underlying dynamics is a central goal of modern physics. This principle of universality reveals that disparate systems can exhibit identical limiting behaviors, irrespective of their microscopic intricacies. For instance, in condensed matter, different physical systems can exhibit identical critical exponents near a phase transition, determined only by fundamental symmetries rather than microscopic details~\cite{Goldenfeld1992,Chaikin1995}. Similarly, in fluid dynamics, the Navier-Stokes equations describe the long-wavelength behavior of materials as different as water and plasma using only a few universal parameters like viscosity and density~\cite{Landau1987,Batchelor1967}.

Here, we demonstrate that this principle of universality provides a powerful framework for understanding the nonlinear quantum optics of subwavelength atomic arrays. These systems, where atoms are spaced closer than their resonant wavelength, have emerged as promising platforms to engineer light-matter interactions and advance various quantum technologies~\cite{bettles2016enhanced,ExponentialAsenjoGarcia2017,shahmoon2017cooperative,Manzoni2018,bekenstein2020quantum,rui2020subradiant,Patti2021,wei2021generation,Santos2022,srakaew2023subwavelength,Shah2024,Rubies-Bigorda2025Deterministic}. In the single-excitation limit, their collective behavior is well understood: they act as tunable mirrors for photons and support long-lived dark spin waves~\cite{bettles2016enhanced,ExponentialAsenjoGarcia2017,shahmoon2017cooperative,Manzoni2018,Zoubi2011}. Together, photons and spin waves serve as the fundamental excitations of the system.

\begin{figure}[t]
\centering
\includegraphics[width=\linewidth]{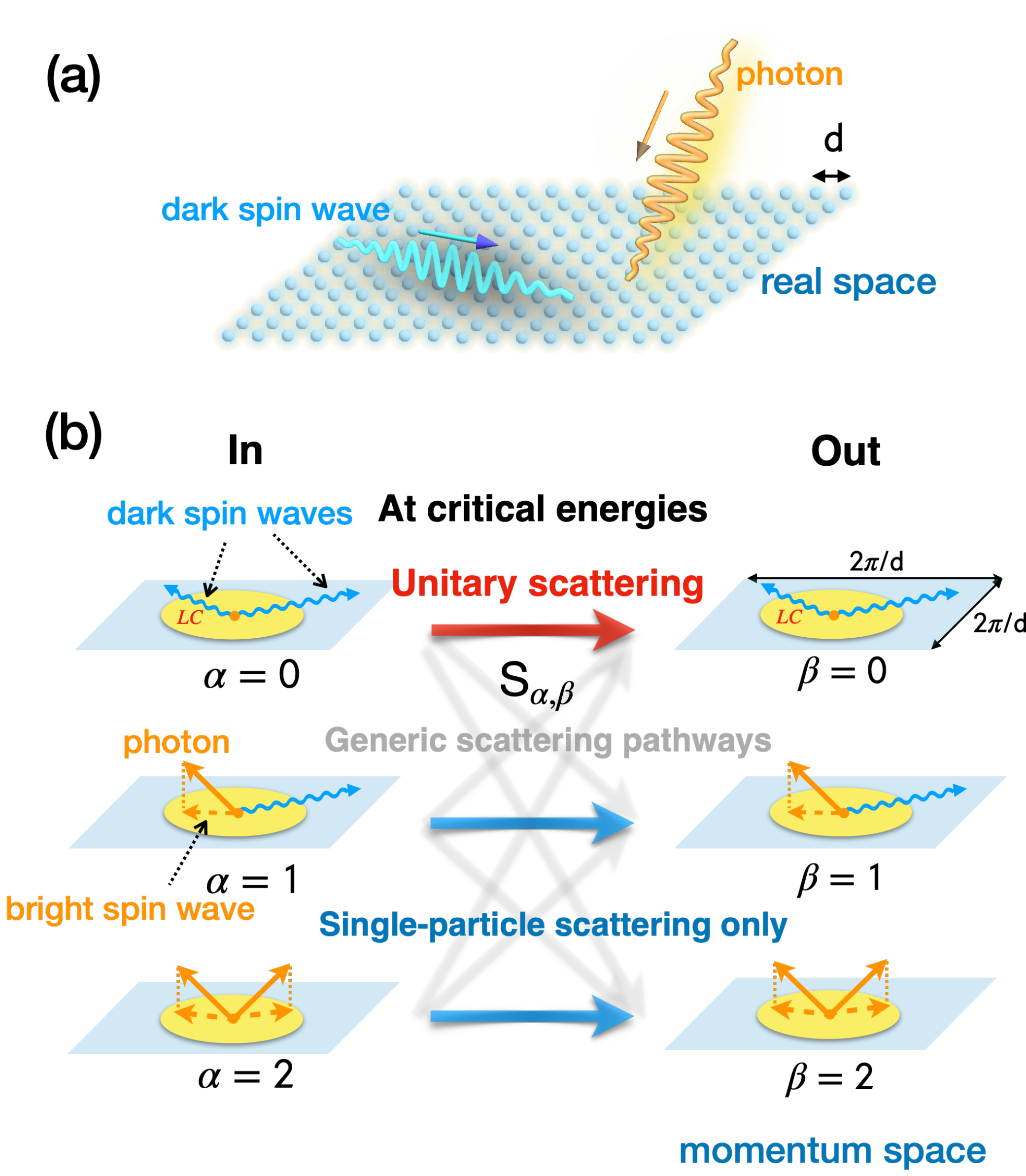}
\caption{(a) Schematic of a subwavelength atomic array in real space, where the lattice spacing $d$ is smaller than half the resonant atomic wavelength ($\lambda_0$). (b) Universal structure of the $S$ matrix at critical energies. The $S$ matrix, $S_{\alpha\beta}$, connects input ($\alpha$) and output ($\beta$) channels defined by their constituent excitations, shown in momentum space. These excitations are photons and dark spin waves. Each photon (orange arrow) has an in-plane momentum corresponding to a bright spin wave (dashed arrow) inside the light cone (LC), while dark spin waves (blue wavy arrows) have momenta outside the LC and are long-lived. Generically, all kinematically allowed channels are coupled (faint gray arrows). At the critical energies, however, the $S$ matrix becomes block-diagonal: the purely atomic channel ($\alpha,\beta=0$) becomes unitary, while all nonlinearities are suppressed in the photonic sectors ($\alpha,\beta>0$), leaving only single-particle dynamics.}
\label{fig:scattering_illu}
\end{figure}

However, harnessing atomic arrays for quantum nonlinear optics requires moving beyond this single-excitation picture. Two-excitation scattering is a crucial step into this nonlinear regime, but it presents a formidable challenge. A complete description must track the many possible scattering processes between incident pairs of photons and dark spin waves (see Fig.~\ref{fig:scattering_illu}), a problem complicated by the mixed dimensionality of the excitations (i.e., interactions between 3D freely propagating photons and 2D lattice-bound spin waves). Consequently, existing approaches often rely on numerical simulations, which are inherently limited to finite systems~\cite{moreno2021quantum,zhang2022photon} or employ various approximations~\cite{cidrim2020photon,bettles2020quantum,williamson2020superatom,Ostermann2024,RubiesBigorda2023,Scarlatella2024,scarlatella2024subwavelength}. While previous analytical approaches have solved the photonic or atomic subspaces~\cite{Solomons2023,Pedersen2024,Zhang2019,Iversen2021,Ke2019}, a holistic description for scattering within and in between the photonic and atomic subspaces is much needed.

In this Letter, we overcome this complexity by discovering a universal structure governing two-excitation scattering. This stems from a unifying principle across scattering theory: at critical points in the dispersion relation, such as scattering thresholds or band saddle points (Van Hove singularities), observables exhibit universal limiting or scaling behaviors~\cite{Wigner1948ThresholdCrossSections,Sadeghpour2000CollisionsNearThreshold,Wojcik2021,Guan2016,wang2022universal,Wang2025UniversalScattering}. Applying this principle to subwavelength atomic arrays, we demonstrate that near these critical points in the two-excitation band structure, the entire $S$ matrix and all partial cross sections collapse into a universal form. This simplification applies across all scattering channels, encompassing every combination of interacting photons and dark spin waves. Crucially, these critical energies are determined by the system's single-excitation properties alone, circumventing the need to solve the full, complex two-body problem. We analytically verify this universal behavior using a comprehensive scattering formalism developed in a parallel work~\cite{Wang2025MultiExcitation}. Building upon recent approaches~\cite{Solomons2023,Pedersen2024}, our framework generalizes the treatment to include all scattering channels (including dark spin waves), enabling the exact calculation of the relevant observables. Ultimately, this discovery offers a powerful predictive shortcut, providing holistic, analytical insights into the rich nonlinear scattering properties of subwavelength atomic arrays.

\textit{System Hamiltonian---}The total Hamiltonian $H=H_{\mathrm{quad}}+U$ comprises a single-excitation part $H_{\mathrm{quad}}$ and a two-excitation interaction $U$. It is derived using the Holstein-Primakoff transformation, as discussed in Refs.~\cite{Pedersen2024,Wang2025MultiExcitation}, which maps the two-level atomic excitations onto a lattice of hard-core bosons. $H_{\mathrm{quad}}$ conserves lattice momentum $\bm{p}$ (the quasimomentum along the array dimension) and decomposes as $H_{\mathrm{quad}}=\int_{\mathrm{BZ}}d\bm{p}\, H_{\bm{p}}$~\cite{MomentumMeasure}, where BZ denotes the atomic Brillouin zone. Each component $H_{\bm{p}}$ governs the dynamics of collective atomic spin waves, created by the bosonic operators $b_{\bm{p}}^\dagger$, and their coupling to the corresponding photon channels. The atomic resonance frequency $\omega_{eg}$ defines the light cone radius $k_0=\omega_{eg}/c$. In the subwavelength regime, where the lattice spacing $d<\lambda_0/2$ (with $\lambda_0=2\pi c/\omega_{eg}$), this light cone is contained entirely within the Brillouin zone, thereby defining two physically distinct regions of momentum space. We work in a rotating frame at the atomic resonance frequency $\omega_{eg}$, effectively setting it as the zero of energy for excitations.

\begin{figure}[t]
\centering
\includegraphics[width=\linewidth]{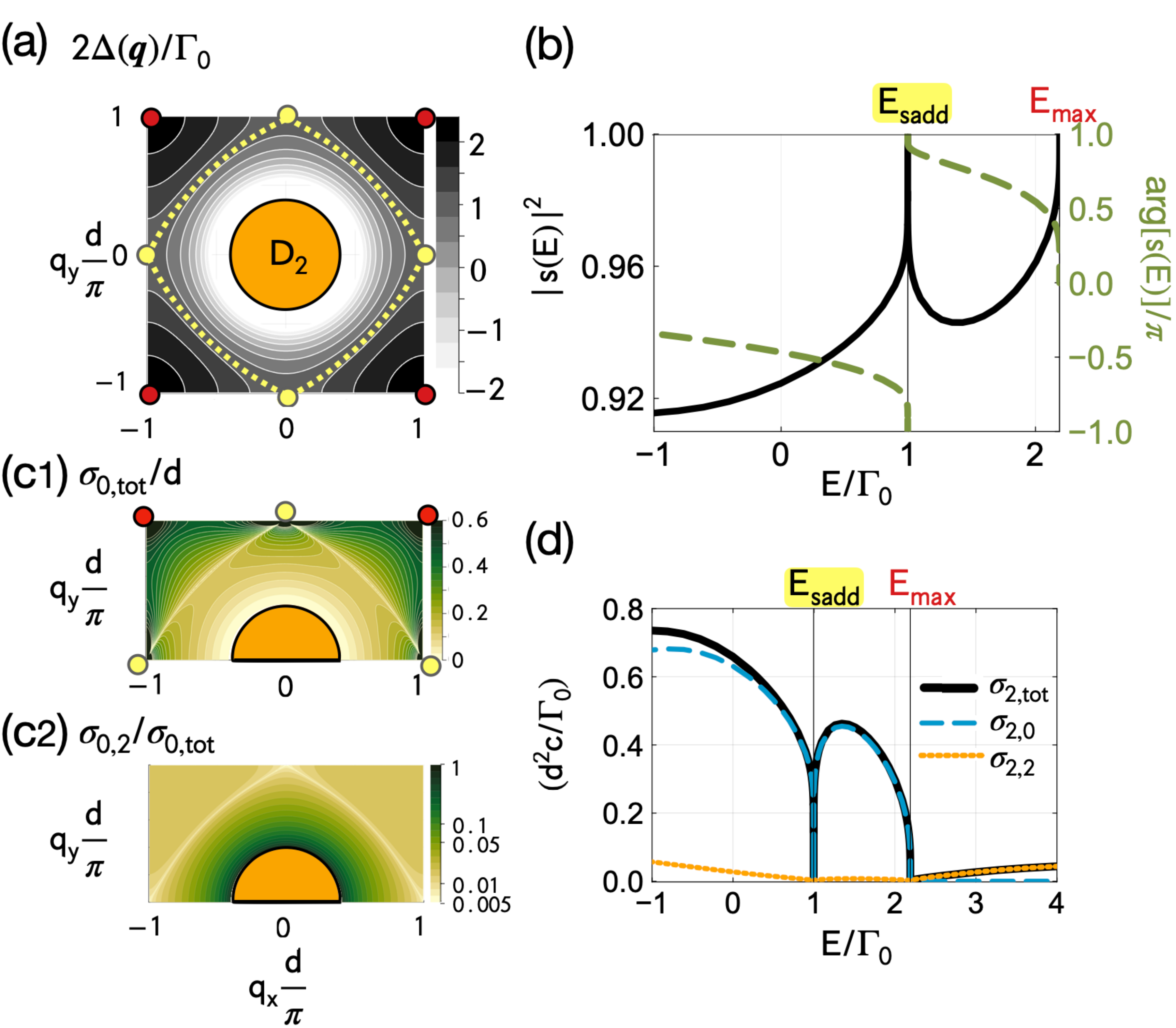}
\caption{Universal scattering signatures near critical points for $\bm P=0$. We consider a square array with lattice spacing $d=0.2\lambda_0$, where the atomic transitions are $\sigma^+$ polarized (dipole moment $\propto\hat{\bm{x}}+i\hat{\bm{y}}$) with respect to the out-of-plane direction $\hat{\bm z}$. (a) The two-dark-wave dispersion relation $\Delta^{(2)}(0,\bm q)=2\Delta(\bm q)$. Red and yellow markers correspond to the band maximum ($E_{\mathrm{max}}$) and saddle points ($E_{\mathrm{sadd}}$), respectively. The orange circle ($D_2$) marks the subspace of two bright spin waves. (b) The nontrivial eigenvalue of the $S$ matrix for two incoming dark spin waves, $s(E)$, as a function of their total energy $E$. In agreement with the universal prediction of Eq.~\eqref{eqSEmax0_final}, the squared magnitude $|s(E)|^2$ (black line) approaches 1 at the critical points, while the phase $\arg[s]$ (green dashed line) approaches 0 (at $E_{\mathrm{max}}$) and $-\pi$ (at $E_{\mathrm{sadd}}$). (c) Cross section for incident dark waves ($\alpha=0$) as a function of their incoming relative momentum $\bm q$. (c1) The total cross section $\sigma_{0,\mathrm{tot}}$ is enhanced at the critical points and suppressed along the saddle lines. (c2) The branching ratio into photons, $\sigma_{0,2}/\sigma_{0,\mathrm{tot}}$, vanishes at the saddle points and along the saddle lines, but remains finite at the maximum, revealing the distinct universal character of the two critical energies. (d) For two normally incident photons ($\alpha=2$), all cross sections vanish at the critical energies, confirming the decoupling of photons and dark spin waves.}
\label{fig:results_all}
\end{figure}

Outside the light cone ($\|\bm p\|>k_0$), the requirement to conserve in-plane momentum implies that the atomic spin waves $b^\dagger_{\bm p}$ are decoupled from propagating photons, as they possess a larger wave number than a resonant free-space photon. These states thus form long-lived dark spin waves with Hamiltonian $H_{\bm p}=\Delta(\bm p)b^\dagger_{\bm p}b_{\bm p}$, where $\Delta(\bm p)$ is the collective Lamb shift obtained by tracing out off-resonant vacuum electromagnetic modes~\cite{ExponentialAsenjoGarcia2017,Wild2020}. Inside the light cone ($\|\bm p\|<k_0$), excitations $b^\dagger_{\bm p}$ couple radiatively to effective 1D photon channels $C_{\bm p}(\kappa)$ and are referred to as bright spin waves~\cite{Wang2025MultiExcitation}. Each channel $C_{\bm p}(\kappa)$ represents photons propagating away from the array with 3D wave vectors $(\bm p,\kappa)$ and $(\bm p,-\kappa)$, sharing the same in-plane momentum $\bm p$ as the spin wave $b_{\bm p}$. Here, $\kappa(>0)$ is the magnitude of the out-of-plane momentum. The Hamiltonian in the light cone reads
\begin{align}
H_{\bm p}={}&\Delta(\bm p)b^\dagger_{\bm p}b_{\bm p}+\int_0^\infty d\kappa\,E_{\bm p}(\kappa) C^\dagger_{\bm p}(\kappa)C_{\bm p}(\kappa)\nonumber\\
&+\int_0^\infty d\kappa\left[g_{\bm p}C^\dagger_{\bm p}(\kappa)b_{\bm p}+\mathrm{H.c.}\right]\quad(\text{for }\|\bm p\|<k_0),
\label{eq:Hp_bright_rev}
\end{align}
with photon energy $E_{\bm p}(\kappa)=c\sqrt{\|\bm p\|^2+\kappa^2}-\omega_{eg}$. The dispersion $\Delta(\bm p)$ for lattice constant $d=0.2\lambda_0$ is shown in Fig.~\ref{fig:results_all}(a). The photon--spin-wave coupling $g_{\bm p}$ in the effective 1D description is defined to yield the radiative decay rate $\Gamma(\bm p)$~\cite{AsenjoGarcia2017Radiation} of the bright spin wave with momentum $\bm p$. In particular, $g_{\bm p}=\sqrt{\Gamma(\bm p)v_{\mathrm{ph}}/(2\pi)}$, where $v_{\mathrm{ph}}=c\sqrt{1-|\bm p|^2/k_0^2}$ is the photon group velocity evaluated on resonance.

The effective interaction $U$, resulting from the two-level nonlinearity of atoms, conserves total momentum $\bm P$ while scattering relative momentum $\bm q\to\bm q'$~\cite{Pedersen2024,Fauseweh2014,Streib2015},
\begin{equation}
U=\frac{u}{2}\int d\bm q\,d\bm q'\int d\bm P\,
 b^\dagger_{\bm P/2+\bm q'}b^\dagger_{\bm P/2-\bm q'}b_{\bm P/2-\bm q}b_{\bm P/2+\bm q}.
\label{eq:U_interaction_revised}
\end{equation}
Here, $u\to\infty$ represents the infinite energy penalty for double occupancy of two atomic excitations.

\textit{Two-excitation scattering---}We focus on the two-excitation manifold where scattering involves pairs of photons ($C^\dagger_{\bm p}(\kappa)$, $\|\bm p\|<k_0$) and/or dark spin waves ($b^\dagger_{\bm p}$, $\|\bm p\|>k_0$), which serve as the stable asymptotic states of the scattering process. Bright spin waves ($b^\dagger_{\bm p}$, $\|\bm p\|<k_0$) are transient states that decay into photons, making them unstable and unsuitable as asymptotic states. Incoming photons can be prepared using laser beams, while dark spin waves can be initialized through techniques such as rapid modulation of the lattice spacing, dynamic light shifts~\cite{Rubies-Bigorda2022}, or multiphoton excitation schemes~\cite{Rusconi2021}.

The full scattering process, illustrated conceptually in Fig.~\ref{fig:scattering_illu}, connects incoming and outgoing states across nine sectors defined by the number of photons and dark waves, and conserves the total lattice momentum $\bm P$ and energy $E$. We label the subspace with $\alpha$ photons as channel $\alpha$, for $\alpha\in\{0,1,2\}$.

Our parallel work~\cite{Wang2025MultiExcitation} demonstrates that the full multichannel scattering process can be constructed from the dynamics of the atomic spin waves (for purely photonic scattering, see Ref.~\cite{Pedersen2024}). Because of the separation of timescales between the atomic resonance and decay rate, the atomic dynamics can be described by an effective Hamiltonian~\cite{ExponentialAsenjoGarcia2017,Lehmberg1970}, where photons are traced out of the system. The scattering of atomic spin waves is thus described by a complex dispersion relation $\epsilon^{(2)}(\bm P,\bm q)\equiv\epsilon(\bm P/2+\bm q)+\epsilon(\bm P/2-\bm q)$. The single-excitation dispersion $\epsilon(\bm p)$ is real for dark spin waves ($\|\bm p\|>k_0$), where $\epsilon(\bm p)=\Delta(\bm p)$, but acquires an imaginary part for bright spin waves ($\|\bm p\|<k_0$), $\epsilon(\bm p)=\Delta(\bm p)-i\Gamma(\bm p)/2$, where $\Gamma(\bm p)$ represents their decay rate into photons~\cite{AsenjoGarcia2017Radiation}. We denote the real part of $\epsilon^{(2)}(\bm P,\bm q)$ as $\Delta^{(2)}(\bm P,\bm q)$.

\textit{Universal behavior in the $S$ matrix---}We first focus on the purely atomic scattering within the subspace of two dark spin waves (channel $\alpha=0$). The corresponding $S$ matrix, $S_0(\bm P,E)$, which acts on the manifold of two-excitation dark spin waves $\mathcal M_0(\bm P,E)$ with the same total momentum $\bm P$ and energy $E$, has a simple structure due to the momentum independence of the interaction. It possesses a single nontrivial eigenvalue $s(\bm P,E)$ with a corresponding normalized eigenvector of the form~\cite{Wang2025MultiExcitation}
\begin{equation}
|\psi(\bm P,E)\rangle=\mathcal N\int_{\mathcal M_0(\bm P,E)}d\bm q\,\frac{|\bm P,\bm q\rangle}{\sqrt{v_g(\bm P,\bm q)}},
\label{eq:psi_definition}
\end{equation}
where $\mathcal N$ is a normalization factor and $v_g(\bm P,\bm q)=|\nabla_{\bm q}\Delta^{(2)}(\bm P,\bm q)|$ is the relative group velocity. The corresponding eigenvalue $s(\bm P,E)$ is given by $s(\bm P,E)=L(\bm P,E-i0)/L(\bm P,E+i0)$ (see End Matter and Ref.~\cite{Wang2025MultiExcitation}). This central quantity, the local propagator $L(\bm P,\omega)$, represents the system's on-site Green's function~\cite{Pedersen2024,Fauseweh2014,Streib2015},
\begin{equation}
L(\bm P,\omega)=\frac{1}{2}\int_{\mathrm{BZ}}d\bm q\,\frac{1}{\omega-\epsilon^{(2)}(\bm P,\bm q)}.
\label{eqLint2D_final}
\end{equation}
Here, the factor $1/2$ avoids double counting of indistinguishable two-excitation states related by exchanging the two bosons, $\bm q\leftrightarrow-\bm q$ (for fixed $\bm P$). Physically, the imaginary and real parts of $L(\bm P,E\pm i0)\equiv\mp iA+B$ ($A>0$) correspond to $\pi$ times the density of states (DOS) of on-shell and off-shell spin-wave pairs, respectively.

Universality emerges as the energy $E$ approaches the critical energies ($E_{\mathrm{crit}}$) of the dark-spin-wave dispersion $\Delta^{(2)}(\bm P,\bm q)$. These critical energies correspond to the relative momentum $\bm q$ where the group velocity vanishes, which generically occur at band maxima ($E_{\mathrm{max}}$) and saddle points ($E_{\mathrm{sadd}}$) [red and yellow markers in Fig.~\ref{fig:results_all}(a), respectively]~\cite{NoBandMinimum}. As the energy approaches these values, the local propagator $L$ diverges logarithmically, but the nature of this divergence is different for both types of critical energies,
\begin{equation}
L(\bm P,E\pm i0)\propto
\begin{cases}
|\log(|E-E_{\mathrm{max}}|)| &\text{as }E\to E_{\mathrm{max}},\\
\mp i|\log(|E-E_{\mathrm{sadd}}|)|&\text{as }E\to E_{\mathrm{sadd}}.
\end{cases}
\label{eqLsaddlelog_final}
\end{equation}
The distinct scaling of $L$ near the two types of critical points is a direct consequence of their differing dispersion topologies. The dispersion is locally equivalent to $-(q_x^2+q_y^2)$ near a maximum but to $q_x^2-q_y^2$ near a saddle point, a difference formally captured by the transformation $q_x\to iq_x$, which introduces the relative factor of $i$ in the resulting integral.

This reversal of whether the real or imaginary part of $L$ diverges leads to distinct universal limits for the $S$ matrix eigenvalue,
\begin{equation}
s(\bm P,E)\to\begin{cases}
1&\text{as }E\to E_{\mathrm{max}},\\
-1&\text{as }E\to E_{\mathrm{sadd}}.
\end{cases}
\label{eqSEmax0_final}
\end{equation}
Evaluating $L(E\pm i0)$ numerically, we plot $s(\bm P=0,E)$ in Fig.~\ref{fig:results_all}(b), which precisely follows this universal prediction near the critical points. At $E\to E_{\mathrm{max}}^-$, $s\to1$, meaning the $S$ matrix becomes the identity operator and the dark waves pass through each other without interaction. At $E\to E_{\mathrm{sadd}}$, $s\to-1$, and the scattering eigenstate acquires a $\pi$ phase shift.

The physical pictures behind these limits are distinct and intuitive. At a band maximum $E_{\mathrm{max}}$, the DOS of on-shell states is finite, but the DOS of off-shell excitations diverges ($\operatorname{Re}[L(E\pm i0)]\to+\infty$). The incoming spin-wave excitations become ``dressed'' by a diverging amount of off-shell excitations, which suppresses their interaction and leads to $s\to1$. Conversely, at a saddle point $E_{\mathrm{sadd}}$, the DOS of available on-shell states diverges ($|\operatorname{Im}[L(E\pm i0)]|\to\infty$). Here, the incoming spin-wave pair encounters an infinite reservoir of available states to scatter into, resulting in a strong scattering event. Driven by this divergence in $\operatorname{Im}[L(E\pm i0)]$ alongside a finite $\operatorname{Re}[L(E\pm i0)]$, the $S$-matrix eigenvalue is forced to $s\to-1$ near the saddle point, yielding a $\pi$ phase shift.

Crucially, the scattering behavior is not uniform across the equienergy contour. At $E=E_{\mathrm{sadd}}$, the Hilbert space effectively fractures: the scattering eigenstate $|\psi(\bm P,E_{\mathrm{sadd}})\rangle$ becomes localized only on the discrete saddle points in momentum space. This means that dark-wave pairs on the ``saddle lines'' connecting these points [yellow dashed lines in Fig.~\ref{fig:results_all}(a)] are orthogonal to the scattering eigenstate and form a separate, noninteracting channel where the $S$ matrix is simply the identity operator.

At the critical energies, the unitarity $|s(\bm P,E)|\to1$ implies that the channel of two dark spin waves becomes completely decoupled from the photon-containing channels [see Fig.~\ref{fig:scattering_illu}(b)]. This decoupling has a far-reaching effect: in all other scattering channels (i.e., for $\alpha,\beta>0$), inter-excitation interactions are completely suppressed, and each excitation evolves independently according to single-excitation dynamics~\cite{Wang2025MultiExcitation}. Intuitively, this decoupling arises because the divergent density of states (on-shell or off-shell) in the spin-wave channel dominates the scattering phase space, effectively freezing out branching to the finite-density photon-containing channels.

\textit{Universal features in scattering cross sections---}The universal $S$ matrix behavior has interesting and nontrivial implications for the corresponding scattering cross sections. When two plane-wave excitations in channel $\alpha$ impinge on each other with flux $R$ in their center of mass frame, $\sigma_{\alpha,\beta}R$ gives the number of pairs of excitations scattered into channel $\beta$ per unit time. Denoting a specific momentum incoming state in channel $\alpha$ by $\bm k^{(\alpha)}$, this partial scattering cross section is given by (see End Matter and Ref.~\cite{Wang2025MultiExcitation}),
\begin{equation}
\sigma_{\alpha,\beta}=\frac{4\pi|a(\bm k^{(\alpha)})|^2}{v_g(\bm k^{(\alpha)})}\,|\bar T(\bm P,E+i0)|^2\,|\operatorname{Im}[L_\beta(\bm P,E+i0)]|.
\label{eqsigma_CS_revised}
\end{equation}
where $\bm P$ is the projection of the total momentum of the incoming state onto the array, and $E$ is the energy of the incoming state. This expression is a product of four distinct factors: (1) a prefactor $|a(\bm k^{(\alpha)})|^2$, representing the conversion rate from the incoming state to the spin-wave modes of the array~\cite{Wang2025MultiExcitation}, which remains finite and does not contribute to the universal scaling; (2) the inverse group velocity of the incoming state, $v_g(\bm k^{(\alpha)})^{-1}$; (3) the density of states $|\operatorname{Im}[L_\beta]|$ of $\beta$ bright spin waves and $2-\beta$ dark spin waves at $(\bm P,E)$; and (4) the core scattering probability $|\bar T(\bm P,E+i0)|^2$, where we introduce the atomic $T$ matrix $\bar T(\bm P,E+i0)=-L(\bm P,E+i0)^{-1}$~\cite{Pedersen2024,Fauseweh2014,Streib2015}. The fact that $|\bar T|^2$ is independent of the incoming and outgoing momentum is a direct consequence of the on-site interaction, representing a uniform scattering probability into any available pair of atomic spin-wave states. To define $L_\beta$, we partition the Brillouin zone for relative momentum $\bm q$ into domains $D_\beta$ based on the number of constituent bright spin waves [see Fig.~\ref{fig:results_all}(a)]. Within this formalism, the components $|\operatorname{Im}[L_{\beta>0}]|$ represent the density of states for the transient bright-wave modes that serve as the direct gateway to the photonic outgoing states. This analytical transparency allows the universal properties of $L$ to be directly mapped onto all scattering observables, making the origin of the universal phenomena particularly clear.

\begin{figure}[t]
\centering
\includegraphics[width=\linewidth]{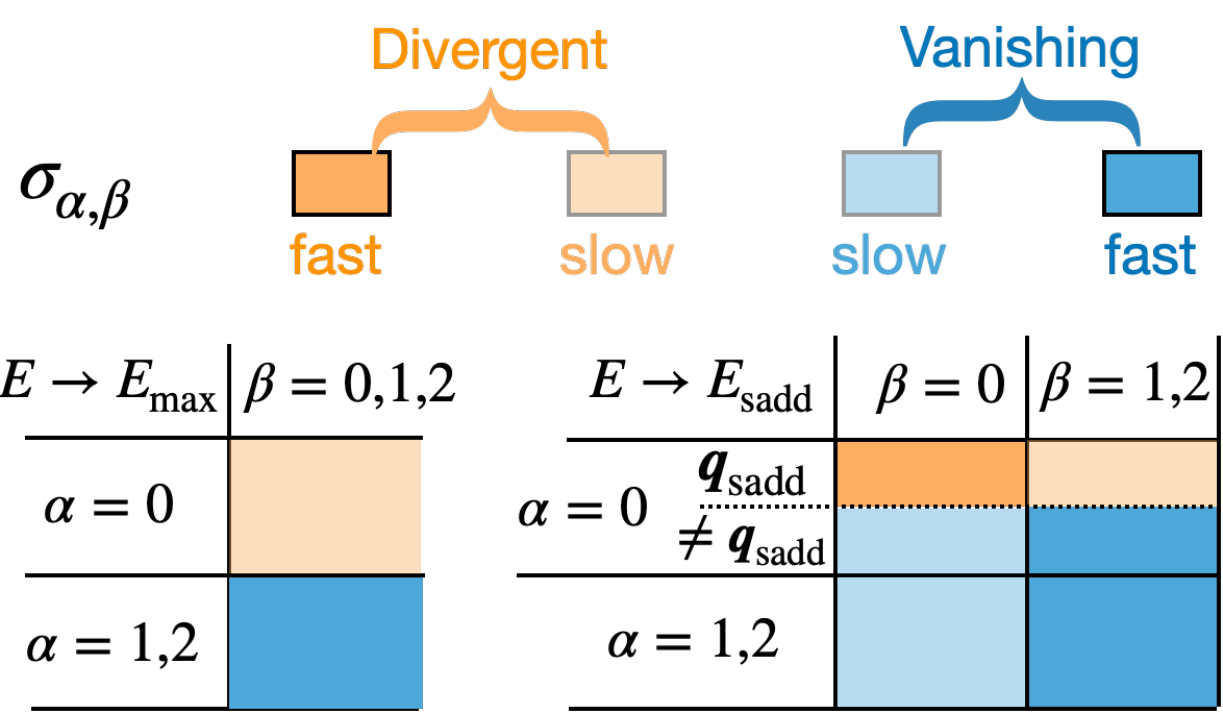}
\caption{Universal scaling of partial cross sections $\sigma_{\alpha,\beta}$ near critical energies ($\Delta E=|E-E_{\mathrm{crit}}|$). The colors denote four scaling behaviors: $\Delta E^{-1/2}|\log(\Delta E)|^{-1}$ (dark orange), $\Delta E^{-1/2}|\log(\Delta E)|^{-2}$ (light orange), $|\log(\Delta E)|^{-1}$ (light blue), and $|\log(\Delta E)|^{-2}$ (dark blue). The tables classify the behavior for each scattering channel $(\alpha,\beta)$ at $E_{\mathrm{max}}$ and $E_{\mathrm{sadd}}$, where for $E_{\mathrm{sadd}}$ the $\alpha=0$ channel is further divided based on whether the incoming relative momentum is at the saddle point ($\bm q_{\mathrm{sadd}}$).}
\label{fig:summary_table}
\end{figure}

For incoming states containing photons ($\alpha=1,2$), the group velocity $v_g$ is finite. As the energy $E$ approaches a critical energy $E_{\mathrm{crit}}$, $\sigma_{\alpha,\beta}$ always vanishes logarithmically, with the specific scaling laws detailed in our summary table, Fig.~\ref{fig:summary_table}. Near band extrema, these scalings reproduce the known 2D Wigner threshold laws~\cite{Sadeghpour2000CollisionsNearThreshold}. This universal suppression is confirmed in Fig.~\ref{fig:results_all}(d) for the case of two normally incident photons.

In stark contrast, for an incoming state of two dark spin waves ($\alpha=0$) with relative momentum $\bm q$ approaching a critical point $\bm q_{\mathrm{crit}}$, the group velocity vanishes as $v_g(\bm k^{(0)})=|\nabla_{\bm q}\Delta^{(2)}(\bm P,\bm q)|\propto\sqrt{\Delta E}$, where $\Delta E=|E-E_{\mathrm{crit}}|$. The diverging inverse group velocity $v_g(\bm k^{(0)})^{-1}\propto\Delta E^{-1/2}$ overpowers the vanishing $|\bar T|^2|\operatorname{Im}[L]|$, leading to a divergent total cross section $\sigma_{0,\mathrm{tot}}=\sum_\beta\sigma_{0,\beta}$.

As also detailed in Fig.~\ref{fig:summary_table}, the specific scaling of this divergence depends on the critical point. The behavior of the total cross section $\sigma_{0,\mathrm{tot}}$ as a function of the incoming relative momentum $\bm q$ is plotted in Fig.~\ref{fig:results_all}(c1). This plot reveals that the cross section is enhanced at the critical points, which are indicated in the figure by the yellow and red markers. Conversely, for incoming states on the saddle lines but away from the saddle points, $v_g$ is finite, and the total cross section is consequently suppressed, as the figure also demonstrates.

The different divergence or vanishing rates of the partial cross sections determine the dominant output channels near the critical energies. We quantify this using the branching ratio into channel $\beta$, $\sigma_{0,\beta}/\sigma_{0,\mathrm{tot}}=|\operatorname{Im}[L_\beta]|/|\operatorname{Im}[L]|$. As $E\to E_{\mathrm{sadd}}$, $|\operatorname{Im}[L]|$ is dominated by the divergent dark-spin-wave contribution $|\operatorname{Im}[L_0]|$, causing the branching ratios into photon-containing channels to vanish. This occurs both at saddle points and along saddle lines away from them, despite the different behavior of $\sigma_{0,\mathrm{tot}}$. By contrast, as $E\to E_{\mathrm{max}}$, all partial densities of states remain finite, so the branching ratios approach finite nonzero limits. Figure~\ref{fig:results_all}(c2) confirms these distinct behaviors.

Finally, we resolve an apparent paradox at $E_{\mathrm{max}}$: the $S$ matrix becomes trivial ($s\to1$), while the cross section $\sigma_{0,\mathrm{tot}}$ diverges. Near the band edge, the trivial $S$ matrix implies that the scattering probability per unit time per excitation pair vanishes, but the vanishing group velocity ($v_g\to0$) of the incoming excitations means the density of excitations within the scattering region increases (when keeping the incident flux constant). The long interaction time afforded by the slow-moving excitations enhances the total cross section, which is the rate of excitations scattered nonlinearly per unit incoming flux. In other words, the diverging cross sections reflect a subtle interplay in which the scattering probability vanishes more slowly than the interaction time grows.

\textit{Discussion---}By fixing the scattering behavior at the critical energies, the universal physics discovered provides a qualitative description of the scattering behavior across a wide energy range, as illustrated in Fig.~\ref{fig:results_all}(d). The prominent suppression of the scattering cross section is anchored precisely at the theoretically predicted $E_{\mathrm{sadd}}$ and $E_{\mathrm{max}}$, while the overall width of the interaction region is set by the natural dissipative scale of the system, $\Gamma(\bm p=0)$ (here, $\approx6\Gamma_0$, where $\Gamma_0$ is the single-atom decay rate). Beyond these sharp features, our results also reveal another generic property: the predominance of scattering into the dark-wave channel ($\beta=0$). This originates from the high density of states associated with the relatively flat dark-wave bands, which enhances the $\sigma_{\alpha,0}$ cross section over other channels. This preferential scattering into subradiant modes is therefore a general feature of atomic arrays supporting flat collective bands~\cite{jaworowski2025laughlin}.

Experimentally, two-photon cross sections can be extracted from the linear dependence of photon transmission or reflection on input intensity~\cite{Wang2025MultiExcitation}. The main experimental test is to measure this cross section as a function of energy near the critical points. Although broadening due to laser resolution, finite size~\cite{FiniteSizeResolution}, and disorder will smooth out sharp theoretical features, a suppression at the critical energies is expected to survive. Observing this residual suppression would validate our predictions within current experimental capabilities. When applied to atomic arrays with strong interactions, such as those utilizing Rydberg-electromagnetically induced transparency schemes~\cite{moreno2021quantum}, this suppression mechanism could serve as an energy-selective switch for two-photon gates, toggling the system between nonlinear and linear behavior.

Although we demonstrate these phenomena using a square array with two-level nonlinearity, the core mechanism is far more general. The universality arises from the local singularity of the phase space near a band-structure critical point, rendering the details of the short-range interaction and lattice geometry irrelevant. While the number and location of these critical points depend on the total momentum $\bm P$ and the specific lattice structure, the scattering physics governed by these band-structure singularities applies to a wide range of nonlinearities and geometries~\cite{GenericMorsePoints}.

The decoupling of dark spin waves from photon channels at critical energies indicates two-excitation subradiant states, discussed further in~\cite{Wang2025MultiExcitation}. This decoupling mechanism provides a theoretical pathway to high-fidelity quantum memories beyond the single-photon limit. This detailed two-excitation scattering analysis underpins future studies of many-body dynamics in these driven-dissipative systems.

\begin{acknowledgments}
We thank Alec Douglas, Simon Hollerith, Sayed Ali Akbar Ghorashi, and Thomas Pohl for helpful discussions. Y.~W. and S.~Y. acknowledge support from the NSF through CUA PFC (PHY-2317134), PHY-2207972, the AFOSR (FA9550-24-1-0311), and the Q-SEnSE QLCI (OMA-2016244). O.~R.-B. acknowledges support from Fundaci\'on Mauricio y Carlota Botton and from Fundaci\'o Bancaria ``la Caixa'' (LCF/BQ/AA18/11680093). V.~W. acknowledges support from the NSF through Grant No. PHYS-2409630 for early-stage investigations of the scattering formalism, and support from the U.S. Department of Energy, Office of Science, Office of Basic Energy Sciences Energy Frontier Research Centers program under Award No. DE-SC0025620 for the subsequent analysis of the dynamics near critical points.
\end{acknowledgments}

\paragraph*{Data availability---}No external datasets were used in this work. The numerical data shown in the figures were generated by evaluating the analytical expressions and integrals described in the manuscript, using the parameters specified there. The data underlying the figures are available from the authors upon reasonable request.

\bibliography{PRL_final_arxiv_library.bib}

\section*{End Matter}
\setcounter{equation}{0}
\renewcommand{\theequation}{A\arabic{equation}}
\renewcommand{\theHequation}{A\arabic{equation}}
\textit{Appendix A: Scattering of two dark spin waves---}Here, following the hard-core treatment of two-level atoms~\cite{Pedersen2024,Fauseweh2014,Streib2015}, we derive the $S$ matrix in the atomic manifold where both the input and output states are a pair of dark spin waves. We derive its nontrivial eigenvalue $s(\bm P,E)$ and the corresponding eigenvector $|\psi(\bm P,E)\rangle$ defined in Eq.~\eqref{eq:psi_definition} of the main text.

The dynamics of the atomic spin waves in a particular excitation number manifold (i.e., assuming no quantum jumps) is described by the effective Hamiltonian after tracing out all photon modes,
\begin{equation}
H=\int_{\mathrm{BZ}}d\bm p\,\epsilon(\bm p)b^\dagger_{\bm p}b_{\bm p}+U,
\label{eq:app_H}
\end{equation}
where $U$ is the hard-core nonlinearity defined in Eq.~\eqref{eq:U_interaction_revised}.

As discussed in Ref.~\cite{Wang2025MultiExcitation}, the standard scattering formalism for unitary potential scattering can be extended to the non-Hermitian Hamiltonian in Eq.~\eqref{eq:app_H}. Restricted to the manifold of two-excitation dark spin waves $\mathcal M_0(\bm P,E)$ with total momentum $\bm P$ and energy $E$, the $S$ matrix $S_0(\bm P,E)$ is related to the atomic $T$ matrix $\bar T(E+i0;\bm P,\bm q,\bm q')$,
\begin{align}
S_0(\bm P,E)={}&\mathds{1}-2\pi i\int_{\mathcal M_0(\bm P,E)}d\bm q\,d\bm q'\nonumber\\
&\times\frac{\bar T(\bm P,E+i0;\bm q,\bm q')}{\sqrt{v_g(\bm P,\bm q)v_g(\bm P,\bm q')}}\nonumber\\
&\times|\bm P,\bm q'\rangle\langle\bm P,\bm q|,
\label{eq:app_S}
\end{align}
where $\mathbb{1}$ is the identity operator in $\mathcal M_0(\bm P,E)$, and $v_g(\bm P,\bm q)$ is the relative group velocity of the dark spin-wave pair $|\bm P,\bm q\rangle$, defined below Eq.~\eqref{eq:psi_definition} in the main text. Here, the atomic $T$ matrix is governed by the two-excitation Lippmann-Schwinger equation,
\begin{align}
\bar T(\omega;\bm P,\bm q,\bm q')={}&U(\bm q-\bm q')\nonumber\\
&+\frac{1}{2}\int_{\mathrm{BZ}}d\bm q''\,\frac{U(\bm q''-\bm q')}{\omega-\epsilon^{(2)}(\bm P,\bm q'')}\nonumber\\
&\quad\times\bar T(\bm P,\omega;\bm q,\bm q''),
\label{eq:app_LS}
\end{align}
where $\epsilon^{(2)}(\bm P,\bm q)$ is the dispersion of the two noninteracting spin waves. The relationship between the $S$ matrix and the $T$ matrix mirrors the standard relationship in unitary potential scattering. The only difference from Hermitian systems is that the dispersion $\epsilon^{(2)}(\bm P,\bm q)$ can be complex due to the dissipative nature of the bright spin waves.

For the specific case of two-level atoms, the nonlinear interaction $U(\bm q''-\bm q')=u\to\infty$ is momentum-independent. As a result, the $T$ matrix must also be independent of the relative momenta $\bm q$ and $\bm q'$. This allows for an exact analytical solution to Eq.~\eqref{eq:app_LS}, yielding $\bar T(\bm P,\omega)=\lim_{u\to\infty}u/[1-uL(\bm P,\omega)]=-L(\bm P,\omega)^{-1}$, where $L(\bm P,\omega)$ is the local propagator defined in Eq.~\eqref{eqLint2D_final}.

Next, we demonstrate that the state $|\psi(\bm P,E)\rangle$ defined in Eq.~\eqref{eq:psi_definition} is an eigenstate of $S_0(\bm P,E)$ with eigenvalue $s(\bm P,E)=L(\bm P,E-i0)/L(\bm P,E+i0)$. Substituting $\bar T=-L(\bm P,E+i0)^{-1}$ into Eq.~\eqref{eq:app_S}, the $S$ matrix can be written as the sum of $\mathbb{1}$ with a projector,
\begin{equation}
S_0(\bm P,E)=\mathbb{1}+2\pi i\frac{\mathcal N(\bm P,E)^{-2}}{L(\bm P,E+i0)}|\psi(\bm P,E)\rangle\langle\psi(\bm P,E)|,
\label{eq:app_projector}
\end{equation}
where $\mathcal N(\bm P,E)$ is the normalization factor of $|\psi(\bm P,E)\rangle$. From this projective form, it is evident that $|\psi(\bm P,E)\rangle$ is an eigenvector of $S_0(\bm P,E)$ with the corresponding eigenvalue $s(\bm P,E)=1+2\pi i[\mathcal N(\bm P,E)^{-2}/L(\bm P,E+i0)]$. Furthermore, any state in $\mathcal M_0(\bm P,E)$ that is orthogonal to $|\psi(\bm P,E)\rangle$ is trivially an eigenstate of $S_0(\bm P,E)$ with an eigenvalue of 1.

The normalization factor $\mathcal N(\bm P,E)$ satisfies $\mathcal N(\bm P,E)^{-2}=\int_{\mathcal M_0(\bm P,E)}d\bm q\,|v_g(\bm P,\bm q)|^{-1}$, which is the DOS of the dark spin-wave pairs at energy $E$ and total momentum $\bm P$. From the definition of the local propagator, this DOS is related to the branch cut discontinuity by $2\pi i\mathcal N(\bm P,E)^{-2}=L(\bm P,E-i0)-L(\bm P,E+i0)$. Substituting this identity into the eigenvalue expression, we obtain the relation $s(\bm P,E)=L(\bm P,E-i0)/L(\bm P,E+i0)$.

\setcounter{equation}{0}
\renewcommand{\theequation}{B\arabic{equation}}
\renewcommand{\theHequation}{B\arabic{equation}}
\textit{Appendix B: Scattering cross-section parameters---}In this appendix, we provide the explicit expressions for the group velocity $v_g(\bm k^{(\alpha)})$ and the term $a(\bm k^{(\alpha)})$ appearing in the partial scattering cross section formula [Eq.~\eqref{eqsigma_CS_revised} in the main text], following the formalism of Ref.~\cite{Wang2025MultiExcitation}.

A two-excitation state in channel $\alpha$ is characterized by its momentum configuration $\bm k^{(\alpha)}$. Depending on whether the excitations are atomic spin waves or photons, we define
\begin{equation}
\bm k^{(\alpha)}=\begin{cases}
\bm p_1,\bm p_2\notin\mathrm{LC}&\alpha=0\ (\text{Dark-Dark}),\\
\bm p_1\in\mathrm{LC},\bm p_2\notin\mathrm{LC}&\alpha=1\ (\text{Photon-Dark}),\\
\bm p_1,\bm p_2\in\mathrm{LC}&\alpha=2\ (\text{Photon-Photon}).
\end{cases}
\label{eq:app_k}
\end{equation}
where for $\alpha=1,2$, the photon states are defined by their transverse wave numbers $\kappa$ (or $\kappa_{1,2}$). For any channel $\alpha$, the state is associated with a total lattice momentum $\bm P=\bm p_1+\bm p_2$ and a relative momentum $\bm q=(\bm p_1-\bm p_2)/2$.

Under the Markov approximation, the group velocity $v_g(\bm k^{(\alpha)})$ associated with state $\bm k^{(\alpha)}$ is given by
\begin{equation}
v_g(\bm k^{(\alpha)})=\begin{cases}
\|\nabla_{\bm q}\Delta^{(2)}(\bm P,\bm q)\|,&\alpha=0,\\
c,&\alpha=1,\\
c\sqrt{2}\sqrt{1-\bm p_1\cdot\bm p_2/k_0^2},&\alpha=2.
\end{cases}
\label{eq:app_v}
\end{equation}
We further define $a_{\bm p}(E)$ as the amplitude for a photon with in-plane momentum $\bm p$ and energy $E$ to excite a single bright spin wave with momentum $\bm p$,
\begin{equation}
a_{\bm p}(E)=\frac{g_{\bm p}}{E-\epsilon(\bm p)}.
\label{eq:app_a}
\end{equation}
The prefactor $|a(\bm k^{(\alpha)})|^2$ in Eq.~\eqref{eqsigma_CS_revised} represents the conversion rate from the incoming state in channel $\alpha$ to the collective spin-wave modes of the array, where the amplitude $a(\bm k^{(\alpha)})$ is given by
\begin{equation}
a(\bm k^{(\alpha)})=\begin{cases}
1,&\alpha=0,\\
a_{\bm p_1}[E_{\bm p_1}(\kappa)],&\alpha=1,\\
a_{\bm p_1}[E_{\bm p_1}(\kappa_1)]a_{\bm p_2}[E_{\bm p_2}(\kappa_2)],&\alpha=2,
\end{cases}
\label{eq:app_amulti}
\end{equation}
where $E_{\bm p}(\kappa)$ is the energy of a single photon with momentum $(\bm p,\kappa)^T$.

\end{document}